\documentclass[twocolumn,amsmath,showpacs,amsfonts,aps,prc,floatfix]{revtex4}
\usepackage{graphicx}
\usepackage{bm}

\newcommand{\la}{\langle}
\newcommand{\ra}{\rangle}
\begin{document}

\title{Effect of charged particle's multiplicity fluctuations on flow harmonics in even-by-event hydrodynamics}
\author{A. K. Chaudhuri}
\email[E-mail:]{akc@veccal.ernet.in}
\affiliation{Theoretical Physics Division\\Variable Energy Cyclotron Centre
\\1/AF, Bidhan Nagar, 
Kolkata 700~064, India}

\begin{abstract}
In nucleon-nucleon collisions, charged particle's multiplicity fluctuates. We have studied the effect of
multiplicity fluctuation on flow harmonics in nucleus-nucleus collision in event-by-event hydrodynamics.
Assuming that the charged particle's multiplicity fluctuations are governed by the negative binomial distribution,   the Monte-Carlo Glauber model of initial condition is generalised to include the  fluctuations. Explicit simulations with the
generalised Monte-Carlo Glauber model initial conditions indicate that the multiplicity fluctuations do not have large effect on  the  flow harmonics.
\end{abstract}

\pacs{47.75.+f, 25.75.-q, 25.75.Ld} 

\date{\today}  

\maketitle

\section{Introduction}
 
In recent years, there is much interest in event-by-event hydrodynamics. Event-by-event hydrodynamics takes into account that  in  nucleus-nucleus collisions, participant nucleon positions fluctuates from  event to event. Effect of
participant nucleon position fluctuations is most prominent on 
the azimuthal distribution of the produced particles. In a non-zero impact parameter collision between two identical nuclei, the collision zone is asymmetric. Multiple collisions transform the initial asymmetry   into momentum anisotropy. Momentum anisotropy is best studied by decomposing it   in a Fourier series, 
 
\begin{equation} \label{eq1}
\frac{dN}{d\phi}=\frac{N}{2\pi}\left [1+ 2\sum_n v_n cos(n\phi-n\psi_n)\right ], n=1,2,3...
\end{equation} 
 
\noindent   $\phi$ is the azimuthal angle of the detected particle and 
$\psi_n$ is the  plane of the symmetry of initial collision zone. For smooth initial matter distribution (obtained from geometric overlap of density distributions of colliding nuclei), plane of symmetry of the collision zone coincides with the reaction plane (the plane containing the impact parameter and the beam axis), 
$\psi_n \equiv \Psi_{RP}, \forall n$. The odd Fourier coefficients are zero by symmetry. However, fluctuations in the positions of the participating nucleons can lead to non-smooth density distribution, which will fluctuate on event-by-event basis.  
The participating nucleons then determine the symmetry plane ($\psi_{PP}$), which fluctuate around the reaction plane \cite{Manly:2005zy}. As a result odd harmonics, which were exactly zero for smoothed initial distribution, can be developed. It has been conjectured that third harmonic $v_3$, which is response of the initial triangularity of the medium, is responsible for the observed structures in two particle correlation in Au+Au collisions \cite{Mishra:2008dm},\cite{Mishra:2007tw},\cite{Takahashi:2009na},\cite{Alver:2010gr},\cite{Alver:2010dn},\cite{Teaney:2010vd}. The ridge structure in $p{\bar p}$ collisions also has a natural explanation if odd harmonic flows develop.  Recently, ALICE collaboration has observed odd harmonic flows    in Pb+Pb collisions \cite{:2011vk}. In most central collisions, the elliptic flow ($v_2$) and triangular flow ($v_3$) are of similar magnitude. In peripheral collisions however, elliptic flow dominates. 

Several authors have simulated Au+Au/Pb+Pb collisions at RHIC/LHC, in event-by-event hydrodynamics  \cite{Schenke:2010rr} \cite{Schenke:2011bn}\cite{Schenke:2012wb}\cite{Gale:2012rq}
\cite{Petersen:2010cw}\cite{Holopainen:2010gz}\cite{Werner:2010aa}\cite{Aguiar:2001ac}
\cite{arXiv:1104.0650}\cite{Bozek:2012fw} \cite{Gardim:2011xv}\cite{Gardim:2012dc}  . Event-by-event hydrodynamics takes into account that participant nucleon positions can fluctuate event-by-event. In general, event-by-event hydrodynamics do not account for the multiplicity fluctuations in NN collisions.
However, it is well known from experiments that in nucleon-nucleon collisions, charged particles multiplicity fluctuates strongly.
Experimental data in limited phase space as well as in full phase space are
  well described by the negative binomial distribution \cite{Alner:1985rj},\cite{Alner:1985zc}.  Recently, in \cite{Dumitru:2012yr}, in a model calculation with Color Glass Condensate (CGC) initial energy density distribution, it was observed that the experimental multiplicity distribution in d+Au collisions at RHIC are better explained if multiplicity fluctuations in NN collisions are included. It was also shown   that various moments of the eccentricity of the collision zone in nucleus-nucleus collisions are affected by the multiplicity fluctuations in NN collisions. Fluctuation dominated moments, e.g. triangularity, increase substantially when the effect of multiplicity fluctuations is included. Effect of multiplicity fluctuations on flow coefficients however is not studied in \cite{Dumitru:2012yr}.  If, higher moments of the eccentricity of the collision zone increase substantially due to multiplicity fluctuations, an increase in flow harmonics is expected.

In the present paper, in a simple model, we incorporate the effect of multiplicity fluctuations in event-by-event hydrodynamics and study its effect  on the flow harmonics. The model is substantially different from that used in \cite{Dumitru:2012yr}. While in
\cite{Dumitru:2012yr}, Color Glass Condensate (CGC) model is used to obtain the initial energy density distribution, in the present model, we have used Monte-Carlo Glauber model
to obtain the initial energy density distribution, event-by-event.
Explicit simulations indicate that multiplicity fluctuations   do not have large effect  
on the flow harmonics.

\section{Monte-Carlo Glauber model with multiplicity fluctuations}

In theoretical simulations of event-by-event hydrodynamics, one generally uses the Monte-Carlo Glauber model to obtain the initial energy density distribution in an event. Details of the Monte-Carlo Glauber model can be found in \cite{Alver:2008aq}.
In a Monte-Carlo Glauber model, according to the density distribution of the colliding nuclei,    two nucleons are randomly chosen. If the transverse separation between them is below a certain distance, they are assumed to interact. Transverse position of the participating nucleons is then known in each event. The positions will fluctuate from event-to-event. If a particular event has $N_{part}$ participants,  participants positions in the transverse plane can be labeled as, $(x_1,y_1), (x_2,y_2)....(x_{N_{part}},y_{N_{part}})$. Energy density distribution in the particular event can be obtained by assuming that    each participant deposit energy $\varepsilon_0$ in the transverse plane,  

\begin{equation}\label{eq2}
\varepsilon(x,y) \approx \varepsilon_0 \sum_{i=1}^{N_{part}}  \delta(x-x_i,y-y_i)
\end{equation}

Fluid dynamical models require continuous density distribution and discrete distribution as in Eq.\ref{eq2} cannot be evolved in a hydrodynamical model. To use in a hydrodynamic model, the discrete density distribution has to be converted into a smooth energy-density distribution. This can be done by smearing the discrete participating nucleon positions by some smoothing function, $\delta(x-x_i,y-y_i) \rightarrow g(x-x_i,y-y_i,\zeta_1,\zeta_2..)$,
$\zeta_i$ being parameters of the smoothing function $g$. 
 
\begin{equation} \label{eq3}
\varepsilon(x,y)=\varepsilon_0 \sum_{i=1}^{N_{part}}  g(x-y,x_i,y-y_i,\zeta_1,\zeta_2....)
\end{equation}

One generally uses a Gaussian smoothing function. However, there can be other choices, e.g. in \cite{RihanHaque:2012wp}, a Woods-Saxon distribution function was used for the smoothing. In the present simulations, we have used a Gaussian distribution 
 
\begin{equation}
g_{gauss}(x-x_i,y-y_i,\sigma) \propto e^{-\frac{{(x-x_i)^2+(y-y_i)^2}}{2\sigma^2}  }, \label{eq6}
\end{equation}
  
\noindent of width $\sigma$=0.5 fm.

As it was mentioned earlier, in experiments, charged particle's multiplicity fluctuates in NN collisions.  
How we incorporate the effect of multiplicity fluctuations within the framework of Monte-Carlo Glauber model? Genesis of Eq.\ref{eq3} gives a clue. It was observed that in nucleus-nucleus collisions, charged particles rapidity density ($dN_{ch}/d\eta$) have two components, (i) a soft component which is proportional to the number of participants $N_{part}$, and (ii) a hard component, which is proportional
to the number of binary collisions $N_{coll}$,

\begin{equation} \label{eq3a}
\frac{dN^{AA}_{ch}}{d\eta}= (1-x) \frac{N_{part}}{2} \frac{dN^{pp}_{ch}}{d\eta}
+x N_{coll} \frac{dN^{pp}_{ch}}{d\eta}
\end{equation}

The hard scattering fraction $x$ is small ($\sim$ 0.1-0.2) in RHIC and LHC energy collisions. If we assume that Eq.\ref{eq3a} is   valid in all the transverse positions $r_\perp$, and further assume that initial transverse energy density is proportional to charged particles multiplicity, one can write the transverse energy density distribution, in nucleus-nucleus as, 

\begin{equation} \label{eq3b}
\varepsilon(x,y)= \varepsilon_0 \left [ (1-x) \frac{N_{part}(x,y)}{2}  
+x N_{coll}(x,y)   \right ]
\end{equation}

A large number of experimental data in $\sqrt{s}_{NN}$=200 GeV Au+Au collisions are well
explained in (smooth) hydrodynamic model with initial energy density as given in Eq.\ref{eq3b}. One understand that the Monte-Carlo Glauber model initial energy density distribution in Eq.\ref{eq3} is essentially a generalisation of Eq.\ref{eq3b} with the hard scattering fraction $x$=0 and $\varepsilon_0$ in Eq.\ref{eq3} is essentially a measure of charged particles multiplicity in NN collisions.

Charged particle's multiplicity fluctuations in NN collisions imply that even in a single event, all the $N_{part}$ participant pairs do not produce the same multiplicity i.e., $\varepsilon_0$ fluctuates.   Considering that in NN collisions,  multiplicity fluctuations follow the  negative binomial distribution, Monte-Carlo Glauber model of initial condition can be generalised to include multiplicity fluctuations as,
 
\begin{equation} \label{eq4}
\varepsilon(x,y)= \mathcal{N} \sum_{i=1}^{N_{part}}   n_i . g(x-y,x_i,y-y_i,\zeta_1,\zeta_2....), 
\end{equation}

In Eq.\ref{eq4}, ($x_i,y_i$) are the transverse positions of the participant nucleons,
$g$ is a smoothing functions with parameters $\zeta_i$.
$n_i$  is a random number following the two parameter negative binomial distribution $P(n:\la  n \ra, k)$.

\begin{equation}\label{eq5}
P(n:\la n \ra, k)=\frac{\Gamma(n +k)}{\Gamma(n+1)\Gamma(k)}\left [\frac{\la n \ra}{\la  n \ra + k}\right ]^n \left [\frac{k}{\la n \ra +k}\right ]^k,
\end{equation}

\noindent   The parameters $\la n \ra$ and $k$ corresponds to the average of the distribution and the   width of the distribution.   $\mathcal{N}$ in Eq.\ref{eq4} is an overall normalising factor. It can be fixed to reproduce experimental observable, e.g. multiplicity or transverse momentum distribution. Presently, for a given $\la n \ra$ and $k$, $\mathcal{N}$ is fixed such that event averaged charged particles multiplicity in 20-30\% collisions reproduces the experimental value.

Negative binomial distribution parameters $\la n \ra$ and $k$ in Eq.\ref{eq5} corresponds to pp collisions multiplicity distributions. However, it is possible that in nucleus-nucleus collisions, pp collision characteristic is changed. For example, in \cite{Dumitru:2012yr}, it was shown that experimental data on charged particles multiplicity distribution in d+Au collisions are better fitted with,

\begin{equation}
k=k_{pp}\cdot min(T_A(r_\perp),T_B(r_\perp)) \sigma_0
\end{equation}

In the following, we have simulated 20-30\% Au+Au collisions at RHIC energy, $\sqrt{s}_{NN}$=200 GeV.
 In $\sqrt{s}_{NN}$=200 GeV Au+Au collisions, average charged particles multiplicity per participant $\frac{1}{.5N_{part}}\frac{dN_{ch}}{d\eta}$ slowly increase with collision centrality. In 20-30\% collision,  $\frac{1}{.5N_{part}}\frac{dN_{ch}}{d\eta}\approx $3.3 \cite{Adler:2004zn}. We then fix $\la  n \ra$=3. Simulation results for a higher average multiplicity $\la  n \ra$=10 will also be shown.  For the width parameter, we have considered two possibilities, which we call model I and II  for $N_{ch}$ fluctuations.

\begin{eqnarray}
{\text Model-I:} k&=&k_{pp}\cdot min(T_A(r_\perp),T_B(r_\perp)) \sigma_0 \label{eq10a}\\
{\text Model-II:} k&=&k_{pp} \label{eq11a}
\end{eqnarray}
 
In $\sqrt{s}$=540 GeV pp collisions, in central rapidity, $k_{pp}\approx 2$ \cite{Alner:1985zc}. $k_{pp}$ decreases logarithmically with energy. We then fix $k_{pp}$=2. The difference between model I and II is the following: in model II, $n_i$ in Eq.\ref{eq4} is a random number following a negative binomial distribution with parameters, $\la n \ra =3$ and $k=k_{pp}=2$. In model I, $n_i$ is still a random number following negative binomial distribution with average $\la n \ra$=3, but now $k$ is not a fixed parameter, rather it depend on the participant nucleon positions. 
Here, I briefly mention the essential difference of the present model with that in
\cite{Dumitru:2012yr}. In \cite{Dumitru:2012yr}, initial energy density was obtained from a color glass condensate model. In CGC model initial conditions, initial energy density is assumed to be proportional to gluon multiplicity,

\begin{eqnarray}
\frac{dN^{A+B\rightarrow g}}{dyd^2r_T}&=&K \frac{N_c}{N_C-1}\int \frac{d^2p_T}{p_T^2}\int^p_T d^2k_T \alpha_x(Q)  \\ 
&& \times \Phi\left (\frac{|{\bf p_T}+{\bf k_T}|}{2},x_1\right ) \Phi\left (\frac{|{\bf p_T}-{\bf k_T}|}{2},x_2 \right )\nonumber
\end{eqnarray}

\noindent where $\Phi(k_T,x)$ is the unintegrated gluon distribution function. Above equation is interpreted as average local multiplicity. In each cell $\nabla^2r_T$, of the transverse plane, the actual multiplicity is a Negative Binomial distribution random variable. In other words, in \cite{Dumitru:2012yr}, in each fluid cell of the transverse plane is associated with fluctuations. In the present model, fluctuations are associated with the participant nucleon only.

\begin{figure}[t]
 \center
 \resizebox{0.35\textwidth}{!}{%
  \includegraphics{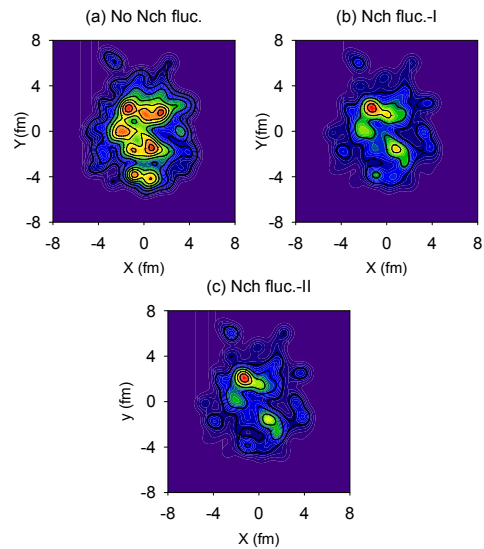} 
}
\caption{(color online) Energy density distribution in a typical Monte-Carlo Glauber model event in 20-30\% Au+Au collisions. (a) without any multiplicity fluctuations, (b)-(c) with multiplicity fluctuations following model-I and II respectively.}
\label{F1}
\end{figure} 

\section{Hydrodynamic equations, equation of state, initial conditions}

With generalised Monte-Carlo Glauber model initial condition for
 initial energy density, space-time evolution of the fluid, in each event  is obtained by solving the energy-momentum conservation equations,

\begin{eqnarray}\label{eq7}
T^{\mu\nu}&=&(\varepsilon+p)u^\mu u^\nu -P g^{\mu\nu}, \\
\partial_\mu T^{\mu\nu}&=&0,
\end{eqnarray}

\noindent where $\varepsilon$ and $p$ are the energy density and pressure respectively, $u$ is the hydrodynamic 4-velocity. We have disregarded any dissipative effect.
Assuming boost-invariance, hydrodynamic equations are solved in $(\tau=\sqrt{t^2-z^2},x,y,\eta_s=\frac{1}{2}\ln\frac{t+z}{t-z})$ coordinate system, with the code AZHYDRO-KOLKATA  \cite{Chaudhuri:2008sj}. Hydrodynamics equations are closed with an equation of state (EoS) $p=p(\varepsilon)$.
Currently, there is consensus that the confinement-deconfinement transition is a cross over. The cross over or the pseudo critical temperature for the  transition  is
$T_c\approx$170 MeV \cite{Aoki:2006we,Aoki:2009sc,Borsanyi:2010cj,Fodor:2010zz}.
In the present study, we use an equation of state where the Wuppertal-Budapest \cite{Aoki:2006we,Borsanyi:2010cj} 
lattice simulations for the deconfined phase is smoothly joined at $T=T_c=174$ MeV, with hadronic resonance gas EoS comprising all the resonances below mass $m_{res}$=2.5 GeV. Details of the EoS can be found in \cite{Roy:2011xt}.

In addition to the initial energy density for which we use the   Monte-Carlo Glauber model,
with or without multiplicity fluctuations,
  solution of hydrodynamic  equations   requires to specify the thermalisation or the initial time $\tau_i$ and fluid velocity ($v_x(x,y),v_y(x,y)$) at the initial time.  A freeze-out prescription is also needed to convert the information about fluid energy density and velocity to invariant particle distribution.  We assume that the fluid is thermalized at $\tau_i$=0.6 fm and the initial fluid velocity is zero, $v_x(x,y)=v_y(x,y)=0$. The freeze-out is fixed at $T_F$=130 MeV. 
We use Cooper-Frye formalism to obtain the invariant particle distribution of $\pi^-$ from the freeze-out surface. Resonance production is included. Considering that pions constitute $\sim$ 20\% of all the charged particles, $\pi^-$ invariant distribution is multipled by the factor  $2\times 1.2$ to approximate the charged particle's invariant distribution. 
From the  invariant distribution, harmonic flow coefficients are obtained as \cite{arXiv:1104.0650},

\begin{figure}[t]
\center
\resizebox{0.35\textwidth}{!}{%
\includegraphics{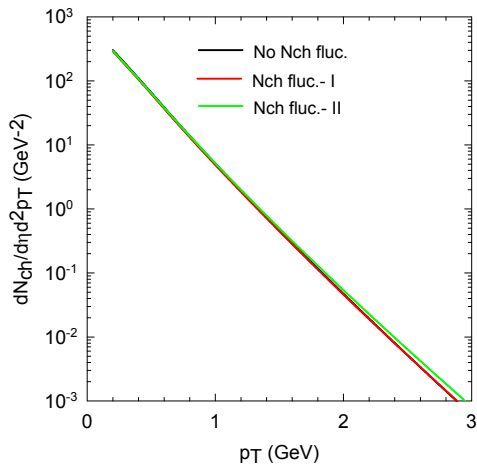} 
}
\caption{(color online) Charged particles transverse momentum distribution with and without    multiplicity fluctuations. }
\label{F2}
\end{figure}

\begin{eqnarray}
v_n(y,p_T)e^{in\psi_n(y,p_T)}&=&\frac{\int d\phi e^{in\phi} \frac{dN}{dy  p_Tdp_T d\phi}}  {\frac{dN}{dy p_Tdp_T}} \label{eq8}\\
  v_n(y)e^{in\psi_n(y)}&=& \frac{ \int p_T dp_T d\phi e^{in\phi} \frac{dN}{dy p_T dp_T d\phi} } { \frac{dN}{dy} } \label{eq9}
\end{eqnarray}
  
In a boost-invariant version of hydrodynamics, flow coefficients are rapidity independent.
Present simulations are suitable only for central rapidity, $y\approx$0, where boost-invariance is most justified. Hereafter, we drop the rapidity dependence. $\psi_n$ in Eqs.\ref{eq8},\ref{eq9} is the participant plane angle for the n-th flow harmonic. 
We characterise the asymmetry of the initial collision zone in terms of various moments of the eccentricity  \cite{Alver:2010gr},\cite{Alver:2010dn},\cite{Teaney:2010vd},

\begin{eqnarray} 
\epsilon_n e^{in\psi_n} &=&-\frac{\int \int \varepsilon(x,y) r^n e^{i2\phi}dxdy}{\int \int\varepsilon(x,y) r^n dxdy} \label{eq11}, n=2,3,4,5
\end{eqnarray} 

$\epsilon_2$ and $\epsilon_3$ are called eccentricity and triangularity. In the following,
$\epsilon_4$ and $\epsilon_5$ will be called rectangularity and penta-angularity respectively. They measure the square-ness and five sided-ness of the initial distribution respectively.  Eq.\ref{eq11} also determine the participant plane angle $\psi_n$.

\section{Results}

To study the effect of multiplicity fluctuations in event-by-event hydrodynamics, we 
have simulated  20-30\%  Au+Au collisions  for $N_{event}$=100 events. 20-30\% centrality corresponds to b=7.4 fm Au+Au collisions. We consider 100 events with impact parameter varying between 7-8 fm. They roughly corresponds to 20-30\% centrality.
Recently in \cite{Chaudhuri:2011qm}, it was shown that with fluctuating initial conditions, event averaged as well as variance of elliptic flow and triangular flow remain approximately unaltered for $N_{event}$=50-2500.  $N_{event}=100$ is then sufficiently large to study the effect of multiplicity fluctuations    in event-by-event hydrodynamics. In each event, Monte-Carlo Glauber model participant positions are smoothed with a  Gaussian function of width $\sigma$=0.5 fm. Multiplicity fluctuations are accounted either with model I or model II. We also simulate events without any multiplicity fluctuations. The overall normalising factor
$\mathcal{N}$ is fixed such that event averaged charged particles multiplicity reproduces the experimental value, $\frac{dN_{ch}}{d\eta}\approx276$ in 20-30\% Au+Au collisions.

 \begin{figure}[t]
\center
\resizebox{0.4\textwidth}{!}{%
\includegraphics{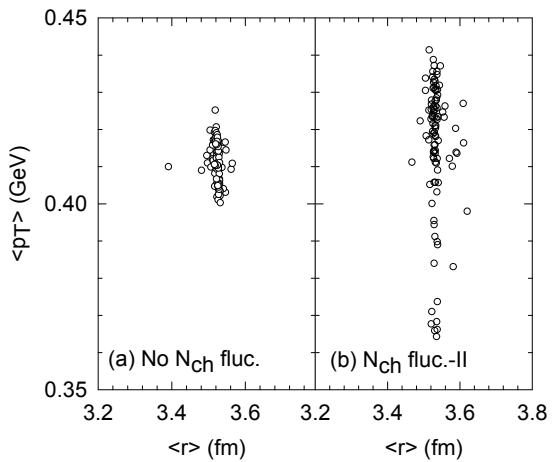}
}
\caption{ (a) average $p_T$ is plotted against the average transverse size in the 100 Monte-Carlo events, when multiplicity fluctuations are neglected. (b) same results when multiplicity fluctuations are included following model-II. }
\label{F3}
\end{figure} 

 \begin{figure}[t]
\center
\resizebox{0.4\textwidth}{!}{%
\includegraphics{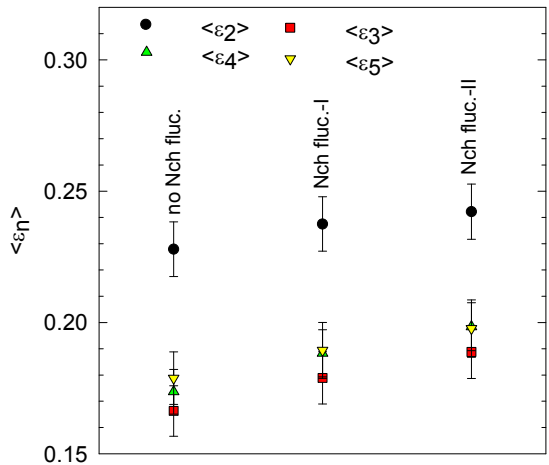}
}
\caption{(color online) Simulation results for event averaged asymmetry parameters $\epsilon_n$, n=2-4. $\epsilon_n$ marginally increase when multiplicity fluctuations are included.}
\label{F4}
\end{figure} 

 In Fig.\ref{F1}, effect of  multiplicity fluctuations   on the initial energy density distribution, in a   typical Monte-Carlo event is shown.   Panel (a) shows the energy density distribution without any multiplicity fluctuations. Energy density distributions with multiplicity fluctuations are shown in panel (b) and (c) respectively for model-I and II.   Effect of multiplicity fluctuations is to smooth the energy density  distribution.
The fine structures which are seen in panel (a) get diffused when multiplicity fluctuations are included.

\subsection{Transverse momentum spectra}

 In Fig.\ref{F2}, simulation results for the charged particle's transverse momentum distribution are shown. The   lines represent the event averaged values. The standard errors are small and are not shown here. 
 The black lines  in Fig.\ref{F2}, are the simulated spectra when multiplicity fluctuations are 
 neglected.  The red and green lines   are simulated spectra with multiplicity fluctuations included following model-I and II respectively. Transverse momentum distribution is marginally affected due to multiplicity fluctuations. Indeed, in the figure, one can not distinguish between spectra obtained without multiplicity fluctuations and with multiplicity fluctuations I. For multiplicity fluctuations II, the spectrum is marginally hardened. 
 
We do note that even though on the average, transverse momentum specrum remain essentially unchanged with or without multiplicity fluctuations, $p_T$ fluctuation increase when multiplicity fluctuations. In Fig.\ref{F3}(a) average transverse momentum in each of the 100 simulated events is plotted against the average transverse size ($\la r \ra$) of the collision zone. The multiplicity fluctuations are neglected. The same results, when multiplicity fluctuations are included following model-II are shown in Fig.\ref{F3}(b). In each event,   average $r$ are   obtained as, 

\begin{equation}
\la r \ra =\left [\frac{\int dxdy (x^2+y^2) \varepsilon(x,y)}{ \int dxdy \varepsilon(x,y)} \right ]^{1/2}
\end{equation}
 
When multiplicity fluctuations are included, $p_T$ fluctuations are increased. Even though $p_T$ fluctuations are increased, event averaged $p_T$ remains approximately same. Fluctuations in average transverse size, however is not increased and event averaged mean transverse size also remain approximately same.

 \begin{figure}[t]
\center
\resizebox{0.4\textwidth}{!}{%
\includegraphics{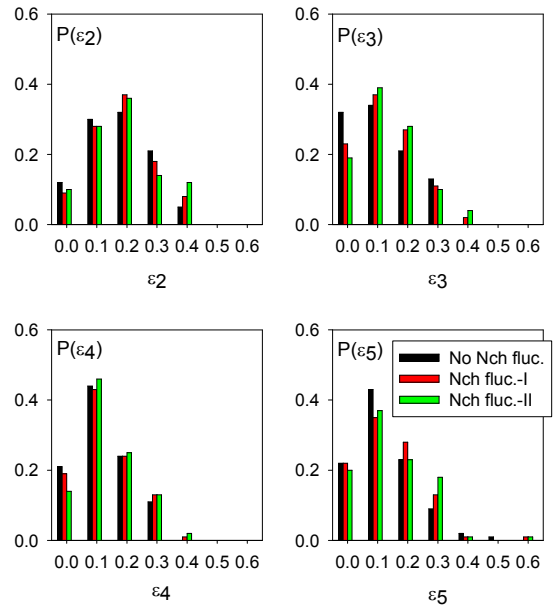}
}
\caption{(color online) Probability distribution of initial spatial anisotropy $P(\epsilon_2)$, $P(\epsilon_3)$, $P(\epsilon_3)$, $P(\epsilon_4)$ and $P(\epsilon_5)$ in simulated events are shown.
In each panel, the black colors represent simulations without multiplicity fluctuations. Multiplicity fluctuations with model I and II are indicated by red and green  colors.}
\label{F5}
\end{figure}

The present result that event averaged transverse momentum distribution remain approximately unaltered even when multiplicity fluctuations are included for, are at odds with results by others \cite{Gale:2012rq}\cite{Bozek:2012fw}. In \cite{Gale:2012rq}
Schenke et al, considered three types of initial conditions, (i)  Glasma initial condition (ii) Monte-Carlo Glauber model initila condition and (iii) Monte-Carlo Color Glass Condensate (KLN) initial condition. 
In addition to fluctuations of nucleon positions,
the Glasma initial condition includes quantum fluctuations of color charges on the length-scale determined by the inverse nuclear saturation scale $Q_s$ and   naturally produces initial energy fluctuations that are described by a negative binomial distribution. In explicit simulations, high $p_T$ particle production is more in Glasma initial condition than in MC-KLN or MC-Glauber initical condition \cite{Gale:2012rq}.
 With Glasma initial condition, hot spots in the initial state are more pronounced than in MC-Glauber or MC-KLN initial condition (see Fig.1 of \cite{Gale:2012rq}), leading to more high $p_T$ production. In the present simulations however, high $p_T$ production is not increased even when multiplicity fluctuations are included in the MC-Glauber model. As shown in Fig.\ref{F1}, with multiplicity fluctuation, fine structures (hot spots) in the initial density distributions get more diffused and high $p_T$ production is not sufficiently increased.
Bozek et al   \cite{Bozek:2012fw} also
studied transverse momentum fluctuations in event-by-event hydrodynamics. It was  shown that for a fixed number of wounded nucleons $N_w$=100, mean $\la p_T \ra$ is anti-correlated with average transverse size.  In the present simulations, the average size of the  fire ball remain essentially unchanged. The average size being determined by the participant nucleon positions, even when fluctuations due to NBD distributions are included, average size do not change much.    $\la p_T \ra$ then remain essentially unchanged, with or without multiplicity fluctuations.

\subsection{Integrated flow coefficients}

In a hydrodynamical model, development of the flow coefficients depend  on the initial spatial anisotropy parameters ($\epsilon_n$)  characterising the collision zone. Approximately, larger the initial asymmetry, more is the flow. Does the initial asymmetry parameters $\epsilon_n$ changes when  multiplicity fluctuations are included in the Monte-Carlo Glauber model of initial condition?
Simulation results for the event averaged asymmetry parameters $\epsilon_n$ and its standard errors are shown in Fig.\ref{F4}. $\la \epsilon_n \ra$ increase marginally when multiplicity  fluctuations are included for. The increase is barely more in model-II than in model I. Marginal effect of multiplicity fluctuations on asymmetry parameter is also evident in Fig.\ref{F5}, where, 
  in four panels, histograms for the probability distribution of the initial eccentricity ($\epsilon_2$), triangularity ($\epsilon_3$), rectangularity ($\epsilon_4$) and penta-angularity ($\epsilon_5$), in the simulated events are shown. The black
bars correspond to simulations without  multiplicity fluctuations. The red and green bars are obtained when  multiplicity fluctuations are included according to model I and II  respectively.   Qualitatively, probability distribution of  asymmetry parameter $\epsilon_n$ remain similar.   The result is not unexpected. In Monte-Carlo Glauber model, the asymmetry parameters are determined mainly by the positions of the participant nucleons (see Eq.\ref{eq11}), weighted by the energy density. Since the participant positions   remain unaltered the change in $\epsilon_n$ due to multiplicity fluctuations is small.   

  \begin{figure}[t]
\center
\resizebox{0.35\textwidth}{!}{%
\includegraphics{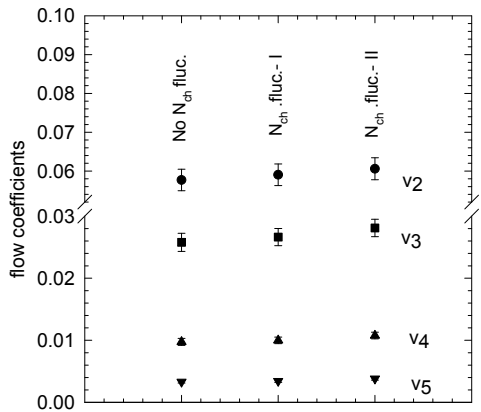}
}
\caption{Simulated flow coefficients $v_n$, n=2,3,4 and 5 with and without  energy fluctuations.}
\label{F6}
\end{figure} 
  
In Fig.\ref{F6}, simulated (integrated) flows, $v_n$, n=2-5, with and without the  multiplicity fluctuations    are shown. The symbols represent the event averaged values, the bars the standard errors. Event averaged flow coefficients marginally increase due to multiplicity fluctuations. However, the increase is small.
In model I of fluctuations, flow coefficients $v_n$, n=2-5  increase by less than 5\%.
In model II also, the increase is modest, less than 5\% in $v_2$ and $v_3$. In higher flow coefficients $v_4$ and $v_5$, the increase is marginally more, $\sim$10-15\%. The result is in agreement with our observations that initial asymmetry parameters depend marginally on the  multiplicity fluctuations in NN collisions.

  \begin{figure}[t]
\center
\resizebox{0.35\textwidth}{!}{%
\includegraphics{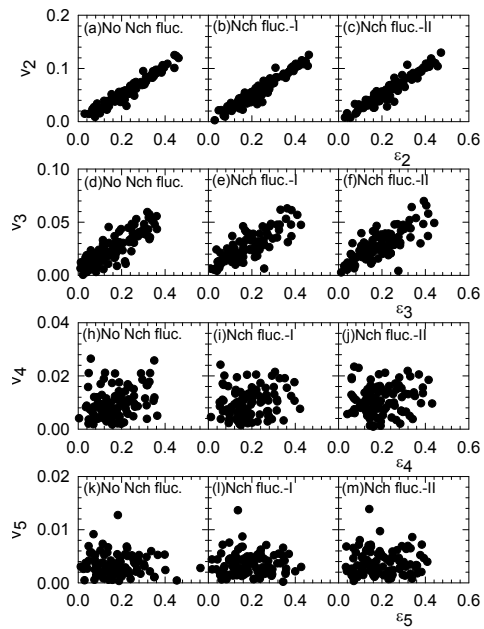} 
}
\vspace{0.4cm} 
\caption{Effect of multiplicity fluctuations on the correlation between flow coefficients and initial asymmetry parameters, ($v_2$,$\epsilon_2$) ($v_3$,$\epsilon_3$), ($v_4$,$\epsilon_4$) and ($v_5$,$\epsilon_5$), are shown.}
\label{F7}
\end{figure} 

In smooth hydrodynamics, elliptic flow ($v_2$) is strongly correlated with initial eccentricity ($\epsilon_2$), more eccentric the initial collision zone, more is the elliptic flow. It is then natural to investigae the correlation between different flow coefficients ($v_n$) and associated asymmetry measures ($\epsilon_n$) in event-by-event hydrodyanmics. Recently, in \cite{Chaudhuri:2011pa}, correlation between elliptic ($v_2$) and triangular ($v_3$) flow with initial eccentricity ($\epsilon_2$) and initial triangularity ($\epsilon_3$) was studied. It was shown that
in event-by-event hydrodynamics, elliptic flow is strongly correlated with initial eccentricity.  Comparatively weak correlation was observed   between triangular flow and initial triangularity. Correlation between the flow coefficients and asymmetry measures in event-by-event hydrodynamic  have been studied also in several other publications  \cite{arXiv:1104.0650},\cite{Gardim:2011xv},\cite{Gardim:2012dc},\cite{Qiu:2012uy}. Similar results are obtained, i.e. higher order flows shows less correlation with the corresponding asymmetry measures. Decorrelation of higher order flows could be understood
as due to nonlinear mixing of modes \cite{Gardim:2011xv}. For example, Gardim et al \cite{Gardim:2011xv}  showed that in order
to correctly predict $v_4$ and $v_5$,  one must take into account nonlinear terms proportional $\epsilon^2$ and $\epsilon_2\epsilon_3$ respectively.
Effect of multiplicity fluctuations on the   correlation between different flow coefficients and asymmetry measures, is also interesting. In Fig.\ref{F7},   simulated flow coefficients $v_n$ are plotted against the initial asymmetry parameters $\epsilon_n$. Simulation results with and without multiplicity fluctuations are shown in separate panels.
One observes that elliptic flow $v_2$ is strongly correlated with initial eccentricity $\epsilon_2$. 
The triangular flow $v_3$ also appear to be correlated with initial triangularity, though the degree of correlation is less than in elliptic flow. However,
 correlation between higher flow coefficients $v_4$ and $v_5$ with spatial asymmetry $\epsilon_4$ and $\epsilon_5$ appears to be weak. Effect of multiplicity fluctuations on the correlation appear to be marginal again.  In \cite{Chaudhuri:2011pa} a quantitative measure was defined to quantify the correlation between flow coefficients and initial spatial asymmetry measure. A modified form is used here to measure the correlation,

\begin{table}[h] 
\caption{\label{table1} Correlation measure for flow harmonics $v_n$, n=2-5 in the event-by-event hydrodynamics with and without multiplicity fluctuations.}
\begin{ruledtabular} 
  \begin{tabular}{|c|c|c|c|c|}\hline
  &\multicolumn{4}{c|}{ $C_{measure}$ } \\ \hline
  & $v_2$ & $v_3$ & $v_4$& $v_5$ \\ \hline 
no Nch fluc. &	$0.980$ & $0.873$ & $0.633$ & $0.712$ \\
Nch fluc.-I& $0.974$ & $0.865$ & $0.538$ & $0.727$\\
Nch fluc.-II& $0.971$ & $0.869$ & $0.466$ &	$0.689$\\  
\end{tabular}\end{ruledtabular}  
\end{table}   

\begin{equation}
C_{measure}(n)=1-\frac{\sum_i [ v_n^i(\epsilon_n) -v_{n,st.line}(\epsilon_n) ]^2}{\sum_i [ v^i_{random}(\epsilon) -v_{st.line}(\epsilon) ]^2}
\end{equation}

$C_{measure}$ 
  essentially measures the dispersion of the simulated flow coefficients from the best fitted straight line, relative to completely random flow coefficients. It varies between 0 and 1. 
If flow coefficients are perfectly correlated then $v_n \propto \epsilon_n$ and $C_{measure}$ is identically unity. For completely random flow coefficients, $C_{measure}$=0. To obtain an even ground for comparison of $C_{measure}$ for different flow coefficients, the flow coefficients ($v_n$) and the asymmetry parameters ($\epsilon_n$) are scaled to vary between 0 and 1.   In table.\ref{table1}, we have listed the $C_{measure}$ values obtained in the present simulations.

Let us first discuss the correlation measures without any multiplicity fluctuations.
Elliptic flow is strongly correlated with initial eccentricity, $C_{measure}(v_2) \sim$ 0.98. Correlation in triangular flow is comparatively weak, $C_{measure} (v_3)\sim 0.87$. 
Correlation between higher flow harmonics, $v_4$ and the initial asymmetry parameter $\epsilon_4$ or between $v_5$ and $\epsilon_5$ is much more weaker than that in elliptic or triangular flow, $C_{measure}(v_4)\sim 0.63$ and $C_{measure}(v_5)\sim 0.71$.
If departure of $C_{measure}$ from unity is a measure of flow uncorrelated with the initial
eccentricity measure, only $\sim$ 2\% of elliptic flow is uncorrelated with initial eccentricity. Uncorrelated flow is more $\sim$13\% for triangular flow. For higher harmonics, $v_4$ and $v_5$, uncorrelated flow is much more, $\sim$ 40\% in $v_4$ and 30\% in $v_5$. Multiplicity fluctuations appear to reduce the correlation between flow coefficients and asymmetry measures. However, with the exception of $v_4$, correlation between flow coefficients and asymmetry measures are reduced only marginally (less than a few percent).  Only in $v_4$, correlation is significantly reduced, e.g. by $\sim$ 15\% in model I of fluctuations and by $\sim$ 26\% in model II of fluctuations.    We conclude that with the exception of flow harmonic $v_4$, the correlation between flow coefficients and initial asymmetry parameter remain largely unaffected by the inclusion of  multiplicity fluctuations in NN collisions. Why the correlation between flow harmonic $v_4$ and the asymmetry measure $\epsilon_4$ is particularly sensitive to multiplicity fluctuations is not understood. More study is required to resolve the issue.

\begin{figure}[t]
\center
\resizebox{0.35\textwidth}{!}{%
\includegraphics{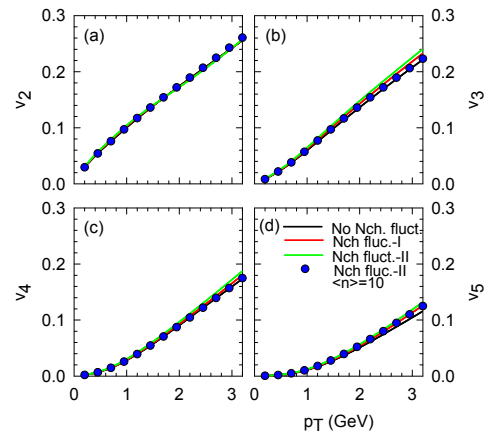} 
}
\caption{(color online) Simulated $v_2$, $v_3$, $v_4$ and $v_5$ are shown in four panels. In each panel, flow coefficients, without multiplicity fluctuations (the black line), with multiplicity fluctuations with model I (the red line) and with model II (the green line) are shown. Average $\la n \ra$ of the negative binomial distribution is fixed at $\la n \ra$=3. The blue circles are simulation results with model II of fluctuations but with $\la n \ra$=10.}
\label{F8}
\end{figure} 

\subsection{Differential flow coefficients}

As it is with the integrated flow coefficients, differential flow coefficients also depend marginally on  multiplicity fluctuations. In Fig.\ref{F8}, in four panels, simulation results for the differential flow coefficients, $v_2(p_T)$, 
$v_3(p_T)$, $v_4(p_T)$ and $v_5(p_T)$ are shown. Elliptic flow is hardly modified by inclusion of fluctuations.
Detailed inspection indicate that in the $p_T$ range 1-2 GeV, differential elliptic flow is decreased by less than $\sim$2\% when  multiplicity fluctuations are included. In triangular flow, flow is marginally increased due to multiplicity fluctuations. In the $p_T$ range 1-2 GeV, in model I of fluctuations, the increase is about $\sim$4-5\%. The increase is marginally more $\sim$8\% in model II. Higher flow coefficients $v_4$ and $v_5$ also increase marginally, e.g. by 6-7\% and by 10-12\% respectively in the $p_T$ range 1-2 GeV. We conclude that differential flow coefficients $v_2(p_T)$-$v_5(p_T)$ are only marginally affected by  multiplicity fluctuations in NN   collisions. 

Negative binomial distribution has two parameters, the width $k$ and average $\la  n \ra$. In the simulations presented here, two possibilities for the width parameter $k$ was  considered.  The average of the distribution was kept fixed at $\la  n \ra$=3. 
To check whether or not simulated flows depend sensitively on the average of the distribution, we have simulated 100 events with $\la  n \ra$=10, $k=k_{pp}=2$.
The results (the blue circles) are shown in Fig.\ref{F8}. Simulated flows with $\la  n \ra$=10 can hardly be distinguished from the simulated flows with $\la  n \ra$=3.

\section{Summary and conclusions}

To summarise, we have studied the effect of multiplicity fluctuations in NN collisions on flow harmonics in $\sqrt{s}_{NN}$=200 GeV Au+Au collisions. The Monte-Carlo-Glauber model for the initial   transverse energy distribution is generalised to include charged particle's multiplicity fluctuations in NN collisions.  Multiplicity fluctuations are assumed to be governed by the negative binomial distribution, with two parameters, the average $\la n \ra$ and the width $k$. 
With the generalised Monte-Carlo-Glauber model initial   transverse energy distribution,
we have simulated 20-30\% Au+Au collisions event-by-event.
  Two possibilities for the width parameter $k$ are considered, (i) model I:
$k=k_{pp}\cdot min(T_A(r_\perp),T_B(r_\perp)) \sigma_0$, i.e. the width depend on the local position of the participant nucleons, and (ii) model II:$k=k_{pp}$, i.e. the width  remain same as in pp collision. In the simulations, we use $k_{pp}$=2. The average the distribution is kept fixed at $\la n \ra$=3. The simulations are constrained to reproduce experimentally charged particles multiplicity in 20-30\% collisions.
 Explicit simulations     indicate that   in event-by-event hydrodynamics, multiplicity  fluctuations do not play significant role  in the development of the flow harmonics. 
Integrated  flows are marginally increased with multiplicity fluctuations. Increase is barely more in model II of fluctuations than in model I. The differential flow is
also increased marginally. Indeed, within the uncertainties integrated and differential flow coefficients  remain largely unaltered irrespective of   multiplicity fluctuations.  We have also studied the correlation between different flow harmonics and initial asymmetry measures. With the exception of fourth flow harmonic ($v_4$), correlation between flow coefficients and asymmetry measures  also remain largely unaltered.  We conclude that in Monte-Carlo Glauber model of initial condition, flow harmonics are not largely affected by the charged particle's multiplicity fluctuations in NN collisions.

\end{document}